\documentclass[11pt,a4paper]{scrartcl}        

\usepackage[utf8]{inputenc} 
\usepackage{amstext} 
\usepackage{amsmath} 
\usepackage{amssymb} 
\usepackage[pdftex,plainpages=false]{hyperref}
\usepackage{verbatim} 
\usepackage{hyperref} 
\usepackage[pdftex]{graphicx}

\newcommand{\eq}[1]{(\ref{eq:#1})}   
\newcommand{\Lb}[1]{\label{eq:#1}}   

\usepackage{tikz}

\newcommand{\cpp}{C$+ +\;$}

\newcommand{\e}{\,\mathrm{e}} 
\renewcommand{\i}{\,\mathrm{i}\,}

\newcommand{\N}{\mathbb{N}}

\newcommand{\R}{\mathbb{R}}     
\newcommand{\C}{\mathbb{C}}

\newcommand{\Hs}{\mathcal{H}}

\newcommand{\ep}{\; .}  

\newcommand{\ec}{\; , \quad} 

\newcommand{\elli}{,\ldots ,}

\newcommand{\id}{\mathbf{1}}

\newcommand{\sclPrd}[2]{\langle \,#1 \,|\, #2 \,\rangle}

\newcommand{\sclPrdSqr}[2]{|\sclPrd{#1}{#2}|^2}
\newcommand{\norm}[1]{\parallel \! #1 \! \parallel}

\newcommand{\map}[2]{#1 \: \rightarrow \: #2 }
   
\newcommand{\set}[2]{\{#1 \; : \; #2\}}  
   
\newcommand{\mapto}[2]{#1~\,~\mapsto~\,~#2 }

\newcommand{\namedmap}[3]{#1 \: : \: #2 \: \rightarrow \: #3 }

\newcommand{\fullmap}[5]{\namedmap{#1}{#2}{#3}  \ec \mapto{#4}{#5}}

\newcommand{\fullmapeqs}[6]
{\begin{equation}\Lb{#1}
\fullmap{#2}{#3}{#4}{#5}{#6}
\end{equation}}

\newcommand{\fullmapeqsi}[7]
{\begin{equation}\Lb{#1}
\fullmap{#2}{#3}{#4}{#5}{#6}#7
\end{equation}}

\begin{document} 

\date{}     
\title{\vspace{0 mm}  On a deterministic disguise of orthodox quantum 
mechanics} 
\author{Ulrich Mutze
\footnote{www.ulrichmutze.de}}\label{fnAuthor}
\maketitle
\begin{abstract}

According to quantum theory, pure physical states correspond to 
equivalence classes of state vectors, where any two members of one class 
differ by a complex factor.
The point is that such a factor does not change the probability for 
the occurrence of any measurement result as computed within the formalism 
of quantum mechanics.
In the formalism to be presented here, the state vector does not only 
determine the probabilities for the occurrence of measurement results, 
rather it determines the result itself.
The information which guides the selection process is obtained from the 
state vector in a way that is not invariant against multiplication by a 
complex factor.
It can be seen as a kind of global phase of the state vectors. 
This global phase then is a perfectly hidden variable.
For this presentation of quantum mechanics we have a remarkable 
parallelism to classical statistical mechanics:
If we would know exactly the initial state of a system, we could compute 
the outcomes for all experiments to be done later on this system. 
In the quantum case the initial state is to be represented by its state 
vector with specified phase, e.g. as a concrete complex-valued wave 
function. 
\end{abstract} 

\newenvironment{Quote}{\begin{list}{}{
                          \setlength{\rightmargin}{\leftmargin}}
                          \item[]``\ignorespaces}
                          {\unskip''
                       \end{list}}

\section{Introduction}
\label{intro}

As is well known, quantum mechanics can be interpreted in a number of 
ways. 
The view on quantum mechanics which underlies the present work is 
similar to the `Copenhagen spirit' as outlined in chapter VII.1 of 
\cite{Haag}. See also \cite{Messiah}, Ch. IV, §§ 15-19.
We thus deal with systems that are amenable to reproducible 
\emph{experiments} consisting of manipulations and observations 
that can be described in an unambiguous 
language.
As N. Bohr put it (cited from \cite{Haag}, p. 295): `We must be 
able to tell our friends what we have done and what we have learned'.

It has tradition 
to characterize this unambiguous language as the language of classical 
physics.
In my eyes it is clearer to characterize it as the language of engineering.
For instance, modern photo-detectors such as CCDs
are based on non-classical concepts of semiconductor physics.
However, as commercially available products, these are the outcome of 
documented manufacturing processes, based on documented designs, guided by 
a documented device theory. 
This device theory is based on authoritative textbooks such as \cite{Sze} 
which don't speak the `quantum language' with states, Hamilton operators, 
and observables.
The existence of an unambiguous language 
is paralleled by the existence of unambiguous actors: electrons, 
photons, atoms, molecules and alike. 
Despite all quantum uncertainty associated with motion and internal 
dynamics of these quantum actors, their constant physical properties like 
rest-mass, charge, and internal angular momentum, together with their 
persistence
(each of the atomic nuclei around us has an individual history 
going back some billion years to its formation in the evolution
of a no longer existing star) 
allow us to perform experiments which deliver reproducible 
statistical data derived from myriads of individual factual measuring 
events.
A typical case of such measuring events are detector `clicks' which say 
where a particle was found.
Assume that the detectors belong to a conventional scattering 
experiment, where a beam of particles penetrates a target foil and the 
scattered particles get registered.
Then it is clear that the detected impacts allow an unambiguous digital 
documentation.
It is also clear that quantum theory gives no recipes for predicting the 
individual impacts.
It is only the probability for finding an impact at an arbitrarily chosen 
position for which quantum mechanics allows to derive formulas.

The idea that the physical content of a motion theory of atomic and 
subatomic particles could be intrinsically statistical is not natural for 
most physicists. 
When the interpretation of Schr\"{o}dinger's $|\psi|^2$ as a 
probability emerged from papers by M. Born and others (see. \cite{Pais}) 
it created disbelief and opposition such as Einstein's (abridged) remark 
`God does not play dice'.
Against Leibniz's `principle of sufficient reason' the impacts should 
happen once here then there and some miraculous principle should tame 
their arbitrariness as to observe the probabilistic rules which are the 
content of the theory.
Never before was a theory proposed of that kind. 
Notice that in classical statistical mechanics of gases each molecular 
collision is determined by the locations and velocities of nearby molecules 
or walls.
The statistical character comes only from the fact that the number of 
dynamical variables is too large for a treatment on the fundamental level. 
With the employment of computers, working on the fundamental level gained 
large new territories~\cite{Verlet}, though.
Determinism and causality as claimed above for collisions allow 
time-evolution to run autonomously.
No authority is required to enforce obedience to the probabilistic rules 
while still respecting conservation laws exactly. 
Rather, the process evolves in the manner of an algorithm.
This does not imply that future states of the process can be predicted; 
strong dependence on initial conditions may make predictions fail. 

These considerations suggest the question whether a logically equivalent 
situation could hold for quantum mechanics as well.
One answer to this question is known: D. Bohm's deterministic 
version of quantum mechanics
\cite{Bohm} augments the Schr\"{o}dinger wave function with particles 
having definite positions and having momenta that are defined in terms of 
the wave function. 
The deterministic time evolution of the wave function 
forces the particles to follow according to a deterministic rule. 
The wave function is said to act as a guiding wave for the particle.
A particle detector `click' is assumed to indicate just the position of 
such a guided particle. Whether a detector will `click' or not thus is 
determined by the `Bohmian state' since it specifies the positions of the 
particles.
Although it is thus a theory with `hidden variables' (the particle 
positions) in a general meaning of the words, it is well understood today
(cf. \cite{Bub}) that it is not covered by von Neumann's no-go theorem on 
hidden variable theories. 
The presently proposed method is literally a hidden 
variable theory as well --- the global phase of the wave function is the 
single hidden variable.
Since it makes exactly the same prediction as orthodox quantum mechanics, 
it is not restricted by \emph{Bell's inequalities} and,
since it does not assume pre-existing values of measured quantities, it 
does not suffer from \emph{quantum contextuality}.

\section{Measurements}
\label{mes}

With respect to mathematical representations of measurements and states we 
follow conventional lines:
A complex Hilbert space $\Hs$, the \emph{normalized} elements of which 
will always be denoted by $\psi$ and referred to as \emph{state vectors} 
or \emph{wave functions}, allows representing (pure) \emph{states} as 
equivalence classes of state vectors with respect to the relation
\begin{equation}\Lb{equivalence}
\psi \sim \psi'\quad \Leftrightarrow \quad \exists\: \omega \in \C_1  
\;\: \psi' = \omega \psi \ec \text{ where}\quad \C_1 := \set{z \in \C}{|z| 
= 1}\ep  
\end{equation}
Wave functions are considered time-dependent according to the 
Schr\"{o}\-dinger picture, where time refers to an inertial system of 
reference which is assumed to be selected once and for all.
This implies that whenever a specific wave function $\psi$ is under 
consideration, there is some 
$t_\psi \in \R$, the point in time for which $\psi$ is valid.

Since all Hilbert spaces of the same dimension are isomorphic, no specific 
features of an experiment are expressed by selecting $\Hs$. 
A connection to the physical world is made by giving certain linear 
operators in $\Hs$ a meaning as descriptors of entities ranging from 
observable quantities to measurement devices, dynamical variables, and 
symmetries.

Measurement devices are represented by self-adjoint operators in $\Hs$. 
Self-adjoint operators viewed as representing measurement devices will be 
called \emph{observables}.

Since measurements always have a limited resolution and a finite 
operational range it is justified to consider as observables only those 
self-adjoint operators which have a spectrum~\footnote{this, of course, 
is the set of possible outcomes of measurements} consisting of 
finitely many points~\footnote{Self-adjoint operators that primarily act 
as generators of unitary one-parameter groups, such as the Hamiltonian, 
are excluded from this restriction.}. 
The general form of an observable $A$ thus is
\begin{equation}\Lb{obsA}
 A = \sum_{a \in \sigma_A} a\, P_a \ec 
\end{equation}
where $\sigma_A$ is a finite subset of $\R$ and $(P_a)_{a \in \sigma_A}$ 
is a family of pairwise orthogonal projectors which add up to the identity 
operator in $\Hs$.
The expression on the right-hand side of equation \eq{obsA} is said to be 
the \emph{spectral representation} of $A$.
Observables in the sense of \eq{obsA} 
fall short of modeling most realistic measurements 
as they are restricted to giving a single number as a result.
This precludes them e.g. from recording the direction of of a particle 
track in space, or, even more so, the tracks of a group of particles.
A sufficient degree of generality is reached if we consider lists 
$(A_1 \elli A_n), n \in \N$, of mutually commuting 
observables.
When interacting with a state vector $\psi$, such a \emph{composite 
observable} 
outputs a list $(a_1 \elli a_n)$ as a result, where $a_i \in \sigma_{A_i}$. 
This list can be considered as the result of a single measurement obtained 
with a complex experimental setup, which may, for instance, comprise some 
spatially separated particle detectors and analyzers.
The $a_i$ may show dependencies, such as $a_1 = - a_2$ due to relations 
between the operators, such as $A_1 = - A_2$, or, less trivially, due to 
algebraic properties of state vector $\psi$ as in the case of the singlet 
state in Section~\ref{compu}.
A composite observable which cannot be extended in a non-trivial 
manner~\footnote{Trivial extensions add self-adjoint operators 
which are functions of the operators which belong the composite observable 
already.}
is said to be a \emph{complete set of commuting observables}. 
This is a standard concept of quantum mechanics, whereas the name 
composite observable is not in common use.

We now turn to constructing a function $\mu$ which takes as arguments an 
arbitrary composite observable $(A_1 \elli A_n)$ and an arbitrary $\psi 
\in \Hs$ and gives as a result a list 
$(a_1 \elli a_n) \in \sigma_{A_1} \times \ldots \times \sigma_{A_n}$.  
In order to simplify indexing we actually consider only the case $ n = 2$ 
as a pattern for $n \in \{ 1, 2, 3, \ldots\}$. Instead of $(A_1,A_2)$ we 
write $(A,B)$, where 
\begin{equation}\Lb{obsB}
 B = \sum_{b \in \sigma_B} b\, Q_b \ep
\end{equation}
Our assumption $A B = B A$ can be shown to be equivalent to $P_a Q_b = Q_b 
P_a$ for all $a \in \sigma_A$ and all $b \in \sigma_B$. 
In such a setting the two spatially separated spin measurements  which are 
considered in D. Bohm's discussion of the EPR-experiment (see e.g. 
\cite{Bohm}, Chapter 7, \cite{Sakurai}, Chapter 3.9 and Chapter 6.3),
and that are the topic of Section \ref{compu}, can be treated as a single 
measurement with a single result (consisting of a pair of values).
That observables corresponding to spatially separated measuring devices 
commute is a foundational principle of local quantum theory \cite{Haag} and 
of quantum field theory \cite{Schlieder}.

Our composite observable $(A,B)$ determines the set $\sigma_A \times 
\sigma_B$ which contains all potential~\footnote{If $B = f(A)$, not all 
$(a,f(a')), a,a'\in \sigma_A$ are possible measurement results, only 
$(a,f(a)), a\in \sigma_A$ are.}  measurement results. 
Which result is to be expected in an actual measurement depends 
on the state vector $\psi$ of the system at the instant of measurement.
Quantum mechanics says that the probability $p(a,b)$
for getting the result $(a,b)$ is given by
\begin{equation}\Lb{prob}
  p(a,b) := \sclPrd{\psi}{P_a Q_b \psi} \ep
\end{equation}
This probability can be written as a  \emph{transition probability}
\begin{equation}\Lb{trans}
p(a,b) = \sclPrdSqr{\psi}{\psi(a,b)} 
\end{equation}
by means of the `collapsed wave function'
\begin{equation}\Lb{final}
\psi(a,b) := \frac{P_a Q_b \psi }{\norm{P_a Q_b \psi}} \ep
\end{equation}
If the system is still existing after a measurement that gave the 
result $(a,b)$ and an immediately subsequent measurement is feasible, then 
the state vector for this measurement has no longer to be taken as $\psi$, 
but as $\psi(a,b)$. This statement is normally referred to as the 
\emph{projection postulate}. 

It is instructive to see that assuming the corresponding relation for 
single observables instead of pairs, allows us to deduce the result for 
pairs and, actually, arbitrarily long lists by successive application: 
\begin{equation}\Lb{chaining}
\psi[b]:= \frac{Q_b \psi }{\norm{Q_b \psi}}\ec
\psi[a,b]:= \frac{P_a \psi[b] }{\norm{P_a \psi[b]}} =
\frac{P_a Q_b \psi }{\norm{Q_b \psi} \frac{\norm{P_a Q_b \psi}}{\norm{Q_b 
\psi}}} = \psi(a,b) \ep
\end{equation}
This approach is, however, less natural than the one we started with. 
This is especially so if our measurement device contains spatially 
separated subsystems $\alpha$ and $\beta$, which are responsible for 
creating the partial results $a$ and $b$ respectively.
Then, we are forced to give the appearance of the partial results a causal 
order. 
Here, in accordance with \eq{chaining}, we let $b$ appear first. 
Then, we need to appeal to the projection postulate and employ the state 
$\psi[b]$ for computing the the probability for $a$ which is consistent 
with \eq{prob} for the pair $(a,b)$ of results.
The state relevant for the computation thus seems to switch from $\psi$ at 
the place of $\beta$ to $\psi[b]$ at the place of $\alpha$.
This, however, is a misconception: the relation between $a$ and $b$ has 
nothing to do with causality as can be seen from the fact that one could
reverse the roles of $a$ and $b$ which then would suggest a causal
connection in the opposite direction.

In conclusion, the physical situation as represented by the measurement 
device $(A,B)$ and the state vector $\psi$ assigns to each element $(a,b) 
\in \sigma_A \times \sigma_B$ a probability $p(a,b)$ as given by \eq{prob}.
The act of measurement selects one of the elements of $\sigma_A \times 
\sigma_B$ in a way which is compatible with these probabilities. Of 
course, not always an element with a major assigned probability will be 
selected; only a probability zero is prohibitive for becoming selected.
The predictive power of the theory for a single measurement is, therefore, 
rather poor.
If, however, the physical situation allows the measurement to be repeated 
with the relevant conditions unchanged, statistics over the selected values 
can be done and the consistency of the obtained measurement data and 
the the probability measure $p$ can be quantified. 

Taking quantum mechanics with M. Born as an inherently statistical theory 
the selection of the actual measurement value out of the manifold of 
possible values is a `true' random event.
Replacing this random event by an algorithmically generated, thus 
deterministic, pseudo-random event brings about the \emph{deterministic 
disguise} announced in the title.

Whereas in biological systems there are many well-understood mechanisms 
that have a (pseudo-)random output (e.g. the distribution of the parent 
genes onto the descendants), it is hard to hypothesize how 
pseudo-random generators in non-living quantum systems could be organized.
Here I assume as a working hypothesis that quantum mechanics (actually all 
Nature) builds on some unknown algorithmic infrastructure~\footnote
{A plausible speculation sees the update rules of a universal discrete 
causal network in this role. 
Since nothing specific from this body of ideas is needed in the present 
work, a single reference: \cite{Wolfram} Chapter 9 and Notes for Chapter 9 
may suffice.
}. 
In the absence of any pertinent knowledge it is a natural assumption that 
the computational strength of this structure is just so that it allows to 
compute all functions which are \emph{computable} in the technical 
sense and that, therefore, a valid pseudo-random number generator 
(RNG) can be realized within this infrastructure (see e.g. \cite{Wolfram} 
Chapter 7, The Intrinsic Generation of Randomness).
We thus take for granted that a RNG is a legitimate building block of a 
physical theory, much like integrals and differential equations.

More specifically, we assume a RNG which outputs uniformly distributed 
real numbers in the interval $[0,1)$. 
The meaning of this is a pragmatic one: the generator passes a 
reasonable set of statistical tests of the hypothesis that the 
output of the generator comes from independent repetition of a stochastic 
experiment which outputs a value $\xi \in [0,1)$ such that the event $\xi 
\in [a,b) \subset [0,1)$ has probability $b-a$. 

A simple example is the RNG \cite{Mutze1}
\begin{equation} \Lb{sf}
   u_i := 1000000 \sin(i) - \lfloor 1000000 \sin(i) \rfloor
\end{equation}
which can be shown to pass all 15 tests of the test battery SmallCrush of 
TestU01 version 1.2.3 (see \cite{Simard}) and fails in 20 of 144 tests 
of the more demanding battery Crush, and in 44 of 160 in the even more 
demanding BigCrush. The latter consumed 16.5 hours of CPU time on a dual 
core 2.4 GHz Laptop Computer.

Forming weighted means out of several RNGs gives again a RNG with 
typically better statistical properties. 
The situation seems to be that for each test battery there can be 
constructed a RNG which passes it completely and for each RNG there can be 
constructed a test battery which lets it fail. 
The real challenge is to identify in this infinite competition a 
David versus Goliath situation, where a relatively simple RNG passes a 
huge, complex, and powerful test battery. 

The value-range $[0,1)$ for RNGs is standard in programming languages and 
comes in naturally in algorithms such as \eq{sf}. In some of the following 
considerations the toroidal value range $\C_1$ would 
be more natural in the first place. The bijection
\fullmapeqs{bij}{\tau}{[0,1)}{\C_1}{\xi}{\exp(2 \pi \i \xi)}
will provide the connection where needed.

In programming languages a random generator is normally a named 
function (such as \verb+rand()+ in the language C) which needs no 
argument to yield a value.
This is, in a sense, a fallacy since the function has a counter as an 
internal state which gets incremented by each function call and which 
works as a hidden argument.
When using pseudo-random generators as legal citizens of theoretical 
physics one should give them a more natural interface, which does not rely 
on a hidden state variable.
Instead we take for granted `pseudo-random clocks'  
\begin{equation}
\Lb{randomclock}
\namedmap{\chi}{\R}{[0,1)} \ec \quad \namedmap{\chi_\tau}{\R}{\C_1} \ec 
\chi_\tau  := \tau \circ \chi \ec   
\end{equation}
which associates a pseudo-random number and a pseudo-random phase with any 
point in time.
Just as numerical values of time are not provided by Nature but by a 
man-made time-keeping system (which has, however, to satisfy conditions 
set by Nature), we consider also the numerical output of the random 
clocks as a part the man-made time-keeping system. 
Actually, on the Internet there are various services which provide both 
pseudo-random numbers and 'true' random numbers (derived from natural 
processes such as radioactivity) on demand.

Pseudo-random numbers can be used to determine rather general 
pseudo-random events by constructing images of the probability space to 
which the pseudo-random numbers belong. 
In our case we construct the mapping 
\begin{equation}
\Lb{rho}
   \namedmap{\rho_\psi}{[0,1)}{\sigma_A \times \sigma_B}
\end{equation}
such that the image of the Lebesgue measure $\lambda$ on $[0,1)$ under 
$\rho_\psi$ is just $p$ from \eq{prob}: 
\begin{equation}
\Lb{image}
\rho_\psi(\lambda) = p \ep  
\end{equation}
The construction of $\rho_\psi$ suggests itself. Define 
\[ X := \{ (a,b) \in \sigma_A \times \sigma_B \:|\: p(a,b) \ne 0 \} \]
and write it in lexicographic order as
\[ X =: \{ x_1 \elli x_n \} \ep \]
Define $\{ y_0, y_1 \elli y_n \}$ iteratively by
$ y_0 := 0,\: y_i := y_{i-1} + p(x_i) $; obviously $y_n = 1$.
Defining the intervals $I_i := [y_{i-1},y_i)$ we have the partition 
\[ [0,1) = I_1 \cup \dots \cup I_n \]
and define $\rho_\psi$ as constant on each of the intervals $I_i$ and 
taking the value $x_i$ on $I_i$.


One way to complete our construction of function $\mu$ would be to require 
each measurement in state $\psi$ at time $t_\psi$ to request the random 
number $\xi := \chi(t_\psi)$ from the random clock and to 
stipulate that the outcome $(a,b)$ of the experiment be given as 
$\rho_\psi(\xi)$.

Since for any $[0,1)$-valued $\lambda$-distributed random variable $x$ the 
$\sigma_A \times \sigma_B$-valued random variable  $\rho_\psi \circ x$ is $p$-distributed
this rule let the measurements, when analyzed statistically, appear 
indistinguishable from measurements that result from the 
intrinsically statistical version of quantum mechanics. 

In this method one may criticize the appearance of $t_\psi$ on grounds of 
the possibility that for many measurements the trigger for measurement 
is not in the device but in the system to be measured (e.g. in the 
collision of two particles). 
Since the formation of the actual measurement values is determined by $\xi$
we need $\xi$ for an unknown time which would destroy the determinism 
of our scheme. 
It is desirable to refer to $t_\psi$ only at the end of a measurement, 
where the measurement result and the point in time exist in documented 
form and belong to the world that can be represented in the `engineering 
language'. 
The method to be proposed in the following achieves this. 

It derives $\xi$ from the state vector $\psi \in \Hs$. 
The simplest choice is $ \xi = \Theta(\psi)$, where $\Theta$ is a function
$\map{\Hs}{\C_1}$ such that $\Theta(\omega \psi) = \omega \Theta(\psi)$ 
for all state vectors $\psi$ and all $\omega \in \C_1$.
There are many such functions and without having input from a 
special situation we have to rely on arbitrary selections. 
Here we choose an orthonormal base $(\phi_i)_{i=1}^{d}, d \in \N \cup 
\{\infty\},$ in $\Hs$ and
set for each state vector $\psi$
\begin{equation}\Lb{kpsi}
k_\psi = \min \set{k \in \N}{\sum_{i=1}^{k} 
|\sclPrd{\phi_i}{\psi}|^2 > 1/2} \ep
\end{equation}
Among the finitely many complex numbers $\sclPrd{\phi_i}{\psi},\: 
1 \le i \le k_\psi,$ we select the one of greatest modulus and name it 
$z$. Then 
\fullmapeqsi{theta}{\Theta}{\Hs}{\C_1}{\psi}{\frac{z}{|z|}}{\ep}
The final formula for the measurement result $(a,b)$ and the desired 
function $\mu$ (see p. 4) is
\begin{equation}
   \Lb{final result}
   (a,b) = \rho_{\psi}(\tau^{-1}(\Theta(\psi))) =: \mu((A,B),\psi) \ep
\end{equation}
In order for this to work we have to manage the circumstances that 
determine the phase of a state vector.
A state comes into existence either by preparation or assumption (typically 
based on a priori knowledge).
Let us consider an experiment dealing with a single isolated hydrogen atom.
The state of it is `self-preparing'; after some time it can be assumed to 
be the ground state. Physicists know the formula 
for the hydrogen atom ground state wave function. No question that 
in Nature there occurs something that corresponds closely to 
the behavior of the mathematical object defined by this formula.
For mathematical deductions concerning future behaviors of the 
object it suggests itself to take the wave function as given by the books 
as initial state vector. 
Actually, any multiple by a $\omega \in \C_1$ would represent the same 
state.
Since in the formalism to be presented here, the overall phase of the wave 
function \emph{does} matter, it would be unnatural to rely here on a 
choice that text book authors made, based on their understanding of 
simplicity of formulas. 
Even in the unmodified quantum mechanics formalism we have to acknowledge 
that the time-dependent Schr\"{o}dinger equation deals with state vectors 
and not with states. 
If we understand time-evolution as an algorithm, as indicated earlier in 
this section, it is clear that this algorithm works on a state vector and 
not on a state.
The logic is similar to an algorithmic implementation of the arithmetic of 
integers: one has to select a base (e.g. 2 or 10) although the results 
depend on this choice in an obvious way which allows to switch from one 
representation to another.
Let this be motivation enough for the following rule:
Whenever a state vector is to be introduced into the description 
of an experiment, there is a `time of birth' $t^*$ for which the 
preparation of the state was finished. Then the `newly born' state vector 
has to be given a phase such that  
\begin{equation}
\Lb{born}
   \Theta(\psi) = \chi_\tau(t^*) \ec
\end{equation}
with $\chi_\tau$ from \eq{randomclock}.

In a computational model of a quantum system formulated in \cpp (and in 
many other object oriented programming languages) an object comes into 
existence through the execution of a constructor function. In such a 
framework one simply has to make this random phase operation \eq{born} the 
last statement in all state constructors.
Also the state vector \eq{final} has to be considered `newly born' and has 
to be `re-phased' as to satisfy
\begin{equation}
   \Lb{rephase}
   \Theta(\psi(a,b)) = \chi_\tau(t) \ec
\end{equation}
where $t$ is the time of completed measurement.

Let us summarize the properties of the proposed modification of orthodox 
quantum mechanics:

\begin{enumerate}
   \item The result of every measurement is strictly determined.
   \item All measurement results occur with the probability predicted by 
quantum theory.
   \item Since each measurement introduces new randomization of the 
phase, there is no way to experimentally find out the mechanism behind 
the built-in determinism.
  \item Determinism is enforced by the same means that a programmer would 
employ in creating a simulation of chained quantum measurements.
\end{enumerate}

\section{A computational low-dimensional model}
\label{compu}

As a proof of concept I implemented the ideas presented so far as an 
interactive free computer program on the web \cite{Mutze2}.
The source code of the program is freely accessible. It is written in 
\emph{Wolfram Language} which is a very expressive universal programming 
language which, due to its freely accessible documentation, can well serve 
as a replacement for expressing algorithmic content in pseudo-code.

The system under consideration is a simplified EPR experiment with two 
electron spins.
By running this program one observes how the simulated measurement results 
approach the quantum mechanical exact result for the spin-spin 
correlation coefficient.
The present form of an article does not allow to present such direct 
evidence.
Nevertheless, a description of the system and the simulated results from a 
single arbitrarily selected run may be a useful illustration of the method 
described in Section \ref{mes}. 

The model is defined by Hilbert space and operators as follows: 

\begin{equation}
\Lb{model}
\begin{split}
\Hs &:= \C^{2} \otimes \C^{2} \ec \\ 
s_x &:= \frac{\sigma_x}{2} \ec s_y := \frac{\sigma_y}{2} \ec s_z := 
\frac{\sigma_z}{2} \\
s_{x}^{1} &:= s_x \otimes \id \ec s_{y}^{1} := s_y \otimes \id \ec
s_{z}^{1} := s_z \otimes \id \ec \\
s_{x}^{2} &:= \id \otimes s_x \ec s_{y}^{2} := \id \otimes s_y \ec 
s_{z}^{2} := \id \otimes s_z \ec \\
S_x &:= s_{x}^{1} + s_{x}^{2} \ec S_y := s_{y}^{1} + s_{y}^{2} \ec
S_z := s_{z}^{1} + s_{z}^{2} \ec \\
S^{2} &:= S_{x}^{2} + S_{y}^{2} + S_{z}^{2} \ec \\
U(\theta_1, \theta_2) &:= \exp(\i \theta_1 s_z) \otimes \exp(\i \theta_2 
s_z) \ec
\end{split}
\end{equation}
where, of course, $\sigma_x, \sigma_y, \sigma_z$ denote the Pauli matrices.
\begin{figure}
\centering
\mbox{
\includegraphics[width=100mm]{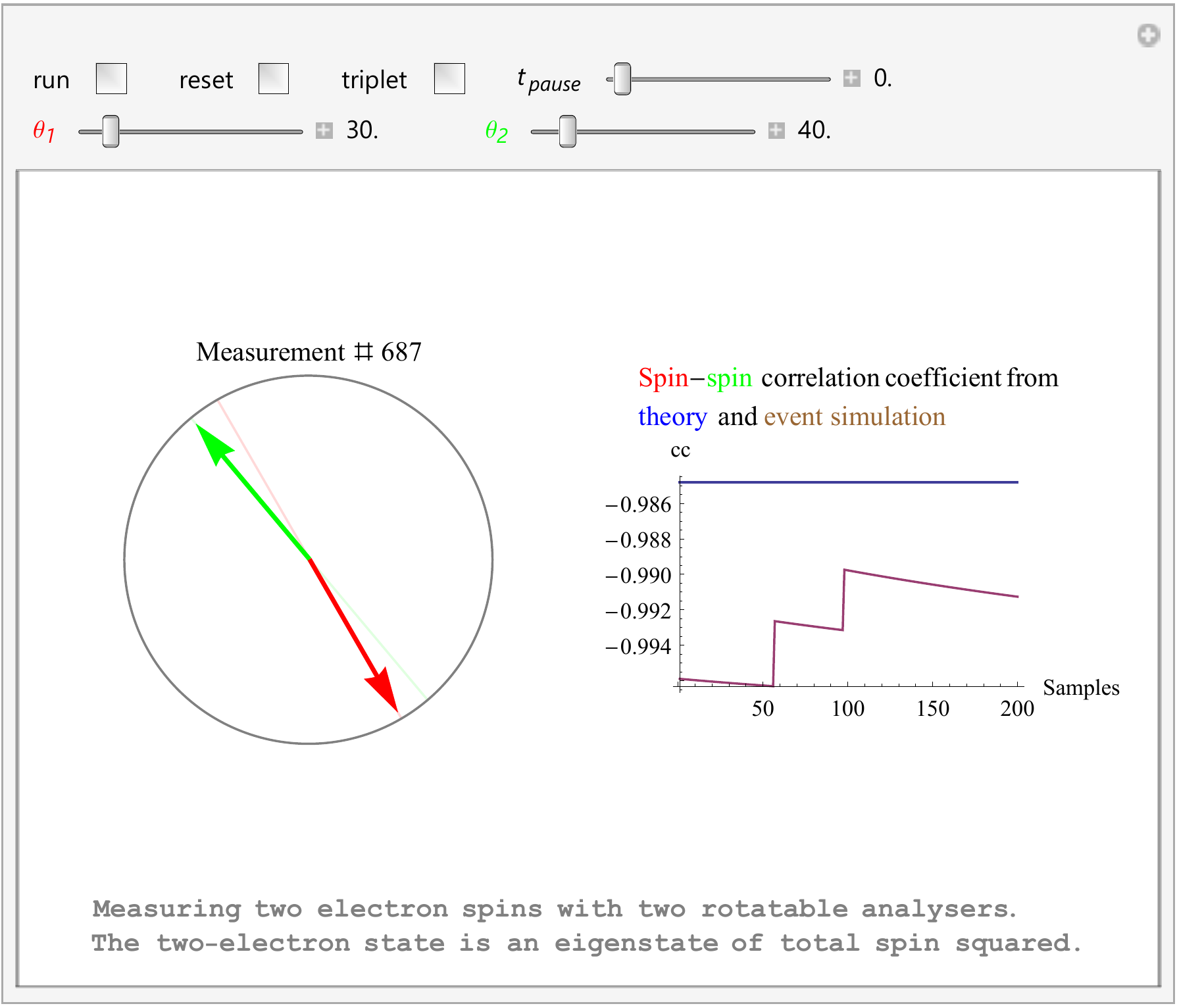}
}
\caption{Simulated measurements of two electron spins along selectable 
directions.}
\label{fig:spinCorrelation}
\end{figure}

All these operators are actually square matrices and their spectral 
representation can be obtained by numerical methods such as the 
function \verb+Eigensystem+ of the computer algebra system Mathematica. 
Of course, for the operator $S^{2}$ we get the eigenvalues $0$ and $2$ 
together with two projectors onto an 1-dimensional and a 3-dimensional 
state space, corresponding to what is known as singlet and triplet states 
respectively. We select a singlet state $\psi$ which is defined uniquely 
up to a phase factor. With the system assumed to be in this state we 
simulate the measurement of the composite observable $(A,B)$ with
\begin{equation}
A := U(\theta_1, \theta_2)\; s_{x}^{1}\; U(-\theta_1, -\theta_2) \ec 
B := U(\theta_1, \theta_2)\; s_{x}^{2}\; U(-\theta_1, -\theta_2) \ep
\end{equation}
When running, the program executes the command~\footnote{See \eq{final 
result} for $\mu$. For the pseudo-random generator the name from Wolfram 
language \cite{WL} is used.}
\begin{equation}
\Lb{mes}
(a,b) = \mu(A,B,\e^{2 \pi \i \text{RandomReal[ ]}}\:\psi)   
\end{equation}
repeatedly, creating a list
\begin{equation}
 (a_1,b_1), (a_2,b_2), (a_3,b_3) \elli (a_n,b_n) \ec  a_i, b_i \in \{ 
1/2, -1/2 \} \ec
\end{equation}
where the repetition stops if some parameter, such as the state $\psi$ or 
the angles $\theta_1, \theta_2$ get changed via the graphical user 
interface. 
The most recent result $(a_n,b_n) = (-1/2, 1/2), n = 687$ is 
shown on the left-hand side of 
Figure~\ref{fig:spinCorrelation} as a red and a green arrow which connect 
the origin $(0,0)$ with points
\begin{equation}
   (a_n \cos (\theta_1 + \frac{\pi}{2}),\: a_n \sin (\theta_1 + 
\frac{\pi}{2}))\:(\text{red}) \ec (b_n 
\cos (\theta_2 + \frac{\pi}{2}),\: b_n \sin (\theta_2 + \frac{\pi}{2}))\: 
(\text{green}) \ep
\end{equation}
The complicating addition of $\pi/2$ is needed since the vertical (and not 
the horizontal) direction was chosen to correspond to $\theta \in 
\{0,\pi\}$. 
In the example shown, the two angle settings differ only by a small angle 
of $10^\circ$ so that there is a probability close to $1$ 
that $a_n$ and $b_n$ differ in sign and, thus, the arrows point in nearly 
opposite directions.

The graphics on the right-hand side deals with the spin-spin correlation 
coefficient~\footnote{We use a simplified definition of a 
correlation coefficient which assumes that the expectation value of the 
quantities to be correlated is zero.}.
The jagged brownish curve shows the most recently obtained 200 values of 
the correlation coefficient of the obtained data: 
\begin{equation}
   c_{n-199}, c_{n-198} \elli c_{n-1}, c_n \ec
\end{equation}
where the $c_i$ are given by 
\begin{equation}
\Lb{cc}
c_i := \frac{\sum_{j=1}^{i} a_j b_j}{\sqrt{\sum_{j=1}^{i} a_{j}^{2}} \: 
\sqrt{\sum_{j=1}^{i} b_{j}^{2}}}  \ep 
\end{equation}
Each of the jags indicates an event with $a_i = b_i$ 
(recall that $a_i = -b_i$ is much more probable).
In all the countless runs I listened, the brownish curve eventually 
reached the theoretical value \eq{correlation}, indicated by the blue 
horizontal line and continued wiggling around this curve. 
Sometimes one observes that rather pronounced deviations from the blue 
line build up. But, as a rule, they have a smaller amplitude as in 
earlier deviation episodes.

The exact quantum mechanical value for the  spin-spin correlation 
coefficient is
\begin{equation}
\Lb{correlation}
\frac{\sclPrd{\psi}{A\,B\, \psi}}{\sqrt{\sclPrd{\psi}{A^2 \,\psi}}\, 
\sqrt{\sclPrd{\psi}{B^2\, \psi}}}  \ep 
\end{equation}
For $\psi$ being the singlet state this quantity can easily be seen to 
equal $-\cos (\theta_1 - \theta_2)$ and thus to yield perfect 
anti-correlation for $\theta_1 = \theta_2$.

\section{Conclusion}
\label{con}
It is rather evident that the method applied here can be generalized as to 
transform any probabilistic theory into a deterministic one. 
The method of converting probabilities into events is, in a sense, a 
reversion of the program of probability theory, which tries to capture the 
characteristics of streams of events by ascribing probabilities to them.

Although we have shown that it does not conflict with logic 
to assume a deterministic mechanism behind the emergence of measurement 
results, one cannot deny that the algorithmic implementation given above 
does not look like a strategy that Nature would follow. 
The algorithmic understanding of physical processes is in its 
infancy at best, as is the understanding of the required resources.

\section{Acknowledgements}
I would like to thank D.P.L. Castrigiano for his critical reading of the 
first version of this article, which led to lots of reformulations and 
additions, and even to the correction of a mathematical error. The 
definition of $\Theta$ contained an infinite sum which was not guaranteed 
to exist. A new definition of $\Theta$ avoids the problem.
I thank Michael J.W. Hall for having made me aware that a construction very
similar to my function $\mu$ is described in \cite{Bell}, end of Chapter IV.

\end{document}